\documentclass[twocolumn,showpacs,preprintnumbers,amsmath,amssymb,prl,aps]{revtex4}

\usepackage{epsfig}% Include figure files
\usepackage{graphicx}% Include figure files
\usepackage{dcolumn}% Align table columns on decimal point
\usepackage{bm}% bold math

\begin{document}

\title{Evolution of a band insulating phase from a correlated metallic phase}

\author{Kalobaran Maiti}
 \altaffiliation{Corresponding author: kbmaiti@tifr.res.in}
\author{Ravi Shankar Singh}
\author{V.R.R. Medicherla}

\affiliation{Department of Condensed Matter Physics and Materials'
Science, Tata Institute of Fundamental Research, Homi Bhabha Road,
Colaba, Mumbai - 400 005, INDIA}

\date{\today}

\begin{abstract}

We investigate the evolution of the electronic structure in
SrRu$_{1-x}$Ti$_x$O$_3$ as a function of $x$ using high resolution
photoemission spectroscopy, where SrRuO$_3$ is a weakly correlated
metal and SrTiO$_3$ is a band insulator. The surface spectra exhibit
a metal-insulator transition at $x$ = 0.5 by opening up a soft gap.
A hard gap appears at higher $x$ values consistent with the
transport properties. In contrast, the bulk spectra reveal a
pseudogap at the Fermi level, and unusual evolution exhibiting an
apparent broadening of the coherent feature and subsequent decrease
in intensity of the lower Hubbard band with the increase in $x$.
Interestingly, the first principle approaches are found to be
sufficient to capture anomalous evolutions at high energy scale.
Analysis of the spectral lineshape indicates strong interplay
between disorder and electron correlation in the electronic
properties of this system.

\end{abstract}


\maketitle

%\section{Introduction}

The investigation of the role of electron correlation in various
electronic properties is a paradigmatic problem in solid state
physics. Numerous experimental and theoretical studies are being
performed on correlated electron systems revealing exotic phenomena
such as high temperature superconductivity, giant magnetoresistance
{\em etc}. Electron correlation essentially localizes the valence
electrons leading the system towards insulating phase. Correlation
induced insulators, known as {\it Mott insulators} are characterized
by a gapped electronic excitations in a system where effective
single particle approaches provide a metallic ground state. The {\it
band insulators} represent insulating phase described within the
single particle approaches. Strikingly, some recent theoretical
studies reveal a correlation induced metallic ground state in a band
insulator using ionic Hubbard model
\cite{andreas,dagotto,mohit,paris}. Such unusual transition has been
observed in two dimensions by tuning effective electron correlation
strength, $U/W$ ($U$ = electron-electron Coulomb repulsion strength,
$W$ = bandwidth) and the local potential, $\Delta$.

In order to realize such effect experimentally, we investigate the
evolution of the electronic structure in SrRu$_{1-x}$Ti$_x$O$_3$ as
a function of $x$, where the end members, SrRuO$_3$ and SrTiO$_3$
are correlated ferromagnetic metal and band insulator, respectively.
Ti remains in tetravalent state in the whole composition range
having no electron in the 3$d$ band\cite{jkim,dd}. Thus, in addition
to the introduction of disorder in the Ru-O sublattice,
Ti-substitution at the Ru-sites dilutes Ru-O-Ru connectivity leading
to a reduction in Ru 4$d$ bandwidth, $W$ and hence, $U/W$ will
increase. Transport measurements\cite{kkim} exhibit plethora of
novel phases such as correlated metal ($x \sim$ 0.0 ), disordered
metal ($x \sim$ 0.3), Anderson insulator ($x \sim$ 0.5), soft
Coulomb gap insulator ($x \sim$ 0.6), disordered correlated
insulator ($x \sim$ 0.8), and band insulator ($x$ = 1.0).

In this study, we have used high resolution photoemission
spectroscopy to probe the density function in the vicinity of the
Fermi level, $\epsilon_F$ and at higher energy scale as well.
Considering the fact that escape depth of the photoelectrons is
small, we have extracted the surface and bulk spectra in every case
by varying the surface sensitivity of the technique. The surface
spectra exhibit signature of disorder at lower $x$ values in
SrRu$_{1-x}$Ti$_x$O$_3$, a metal-insulator transition exhibiting a
soft gap at $\epsilon_F$ for $x$ = 0.5 and a hard gap for higher
$x$. The bulk spectra, on the other hand, reveal an unusual spectral
weight transfer and signature of a {\it pseudogap} at $\epsilon_F$
at higher $x$.

Photoemission measurements were performed using Gammadata Scienta
analyzer, SES2002 and monochromatized photon sources. The energy
resolution for $x$-ray photoemission (XP) and He~{\scriptsize II}
photoemission measurements were set at 300~meV and 4~meV,
respectively. High quality samples of SrRu$_{1-x}$Ti$_x$O$_3$ with
large grain size were prepared following solid state reaction route
using high purity ingredients\cite{sscO1s} followed by a long
sintering (for about 72 hours) at the final preparation temperature.
Sharp $x$-ray diffraction patterns reveal single phase in each
composition with no signature of impurity feature. Magnetic
measurements using a high sensitivity vibrating sample magnetometer
exhibit distinct ferromagnetic transition at each $x$ up to $x$ =
0.6 studied, as also evidenced by the Curie-Weiss fits in the
paramagnetic region. The fits provide an estimation of effective
magnetic moment ($\mu$ = 2.8~$\mu_B$, 2.54~$\mu_B$, 2.45~$\mu_B$,
2.18~$\mu_B$, 2.19~$\mu_B$, 1.95~$\mu_B$ and 1.93~$\mu_B$) and Curie
temperature ($\theta_P$ = 164~K, 156.6~K, 150.6~K, 145.3~K, 139~K,
138.6~K and 100~K) for $x$ = 0.0, 0.15, 0.2, 0.3, 0.4, 0.5 and 0.6,
respectively. The values of $\mu$ and $\theta_P$ for SrRuO$_3$ are
observed to be the largest among those available in the literature
and corresponds to well characterized single crystalline
materials\cite{cao}.

\begin{figure}
\vspace{-2ex}
 \centerline{\epsfysize=4.0in \epsffile{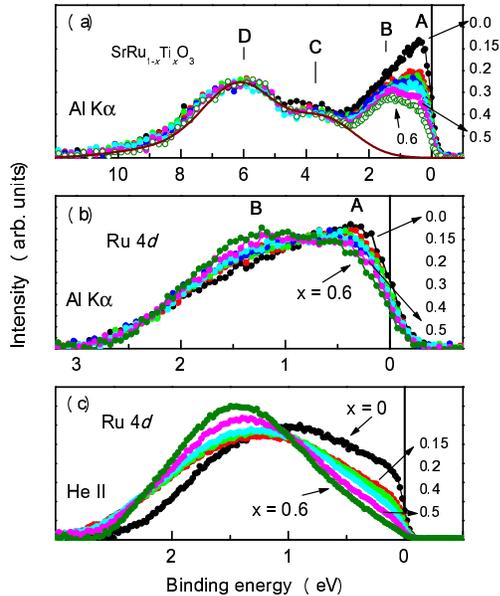}}
%\hspace{2ex}
 \vspace{-10ex}
\caption{(color online) (a) XP valence band spectra of
SrRu$_{1-x}$Ti$_x$O$_3$ for various values of $x$. Solid line
represents the O 2$p$ part for $x$ = 0.6. (b) Ru 4$d$ spectra after
the subtraction of the O 2$p$ contributions as shown in (a). (c) Ru
4$d$ band obtained from He~{\scriptsize II} spectra.}
 \vspace{-4ex}
\end{figure}

In Fig.~1(a), we show the XP valence band spectra exhibiting 4
distinct features marked by A, B, C and D. The features C and D
appear beyond 2.5~eV and have large O 2$p$ character as confirmed
experimentally by changing photoemission cross-sections \cite{ravi}
and theoretically by band structure calculations \cite{kmband}. The
peaks A and B appear primarily due to the photoemission from
electronic states having Ru 4$d$ character. The O 2$p$ part remains
almost the same in the whole composition range as expected. While Ru
4$d$ intensity gradually diminishes with the decrease in
Ru-concentrations, the lineshape of Ru 4$d$ band exhibits
significant redistribution in intensity. In order to bring out the
clarity, we delineate the Ru 4$d$ band by subtracting O 2$p$
contributions. The subtracted spectra, normalized by integrated
intensity under the curve, exhibit two distinct features as evident
in Fig.~1(b). The feature A corresponds to the delocalized
electronic density of states (DOS) observed in {\it ab initio}
results and is termed as {\it coherent feature}. The feature B,
absent in the {\it ab initio} results\cite{kmband}, is often
attributed to the signature of correlation induced localized
electronic states forming the lower Hubbard band and is known as
{\it incoherent feature}. The increase in $x$ leads to a decrease in
intensity of A and subsequently, the intensity of B grows gradually.
Since the bulk sensitivity of valence electrons at 1486.6 eV photon
energy is high ($\sim$ 60\%), the spectral evolution in Fig. 1(b)
manifests primarily the changes in the bulk electronic structure.

\begin{figure}
\vspace{-2ex}
 \centerline{\epsfysize=4.0in \epsffile{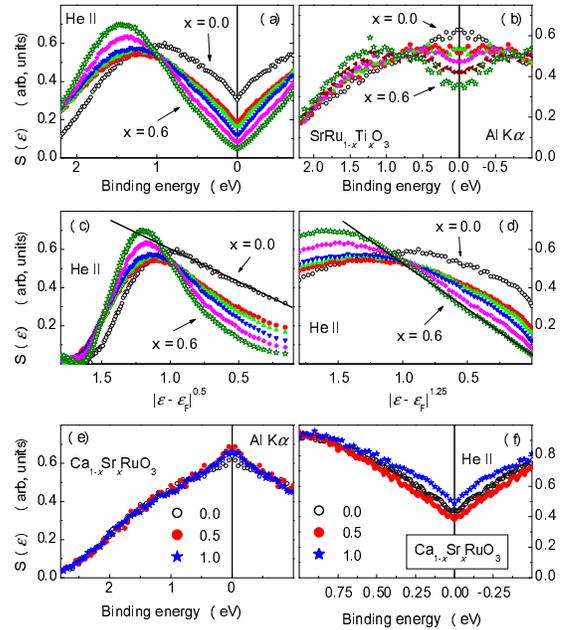}}
%\hspace{2ex}
 \vspace{-8ex}
\caption{(color online) $S(\epsilon)$ obtained from (a) He
{\scriptsize II} and (b) XP spectra of SrRu$_{1-x}$Ti$_x$O$_3$. (b)
$S(\epsilon)$ in (a) are plotted as a function of (c)
$|\epsilon-\epsilon_F|^{0.5}$ (d) and
$|\epsilon-\epsilon_F|^{1.25}$. $S(\epsilon)$ obtained from (e) XP
and (f) He {\scriptsize II} spectra of Ca$_{1-x}$Sr$_x$RuO$_3$.}
 \vspace{-4ex}
\end{figure}

In order to discuss the effect due to the surface electronic
structure, we show the Ru 4$d$ contributions extracted from the
He~{\scriptsize II} spectra in Fig. 1(c), where the surface
sensitivity is about 80\%. Interestingly, all the spectra are
dominated by the peak at higher binding energies ($>$~1 eV)
corresponding to the surface electronic structure as reported in the
case of SrRuO$_3$ and the coherent feature intensity corresponds
essentially to the bulk electronic structure\cite{ravi,fujimori}.
The coherent feature intensity reduces drastically with the increase
in $x$ and becomes almost negligible at $x$ = 0.6. This can be
visualized clearly in the spectral density of states (SDOS) obtained
by symmetrizing ($S(\epsilon) = I(\epsilon) + I(-\epsilon)$;
$I(\epsilon)$ = photoemission spectra, $\epsilon$ = binding energy)
the He {\scriptsize II} and XP spectra. The SDOS corresponding to He
{\scriptsize II} spectrum of SrRuO$_3$ shown in Fig. 2(a) exhibits a
sharp dip at $\epsilon_F$, which increases gradually with the
increase in $x$. The SDOS corresponding to XP spectra in Fig.~2(b),
however, exhibits a peak in SrRuO$_3$ presumably due to large
resolution broadening and intense coherent feature. This peak loses
its intensity and becomes almost flat for $x$ = 0.15 and 0.2.
Further increase in $x$ leads to a {\it pseudogap} at $\epsilon_F$,
which gradually increases with the increase in $x$. Both these
results clearly indicate gradual depletion of SDOS at $\epsilon_F$
with the increase in Ti-substitution.

The effect of resolution broadening of 4 meV in the He {\scriptsize
II} spectra is not significant in the energy scale shown in the
figure. The electron and hole lifetime broadening is also negligible
in the vicinity of $\epsilon_F$. Thus, $S(\epsilon)$ in Fig. 2(a)
provide a good testing ground to investigate evolution of the
spectral lineshape at $\epsilon_F$. The lineshape of $S(\epsilon)$
in Fig. 2(a) exhibits significant modification with the increase in
$x$. We, thus, replot $S(\epsilon)$ as a function of $|\epsilon -
\epsilon_F|^\alpha$ for various values of $\alpha$. Two extremal
cases representing $\alpha$ = 0.5 and 1.25 are shown in Fig. 2(c)
and 2(d), respectively. It is evident that $S(\epsilon)$ of
SrRuO$_3$ exhibit a straight line behavior in Fig. 2(c) suggesting
significant role of disorder in the electronic structure. The
influence of disorder can also be verified by substitutions at the
A-sites in the ABO$_3$ structure. This has been verified by plotting
SDOS obtained from the XP and He {\scriptsize II} spectra of
Ca$_{1-x}$Sr$_x$RuO$_3$ in Fig. 2(e) and 2(f), respectively. Here,
the electronic properties of the end members, SrRuO$_3$ and
CaRuO$_3$ are known to be strongly influenced by the
disorder\cite{kmepl}. Substitution of Sr at the Ca-sites is expected
to enhance the disorder effect. The lineshape of $S(\epsilon)$ in
both Fig. 2(e) and 2(f) remains almost the same across the whole
composition range. Such disorder induced spectral dependence is
consistent with the observations in other
systems\cite{altshuler-aronov,ddsprl} as well.

Interestingly, the lineshape modifies significantly with the
increase in $x$ and becomes 1.25 in the 60\% Ti substituted sample.
Ti substitution introduces defects in the Ru-O network, where
Ti$^{4+}$ having no $d$-electron, does not contribute in the valence
band. Thus, in addition to the disorder effects, the reduced degree
of Ru-O-Ru connectivity leads to a decrease in bandwidth, $W$, which
in turn enhances $U/W$. In systems consisting of localized
electronic states in the vicinity of $\epsilon_F$, a soft Coulomb
gap opens up due to electron-electron Coulomb repulsion; in such a
situation, the ground state is stable with respect to
single-particle excitations, when SDOS is characterized by
$(\epsilon - \epsilon_F)^2$-dependence \cite{efros,massey}. Here,
gradual increase in $\alpha$ with the increase in $x$ in the
intermediate compositions is curious and indicates strong interplay
between correlation effect and disorder in this system.

The extraction of surface and bulk spectra requires both the XP and
He {\scriptsize II} spectra collected at significantly different
surface sensitivities. Thus, we broaden the He {\scriptsize II}
spectra upto 300 meV and extract the surface and bulk spectra
analytically using the same parameters as used before for
CaSrRuO$_3$ system\cite{ravi}. The surface spectra shown in Fig.
3(b) exhibit a gradual decrease in coherent feature intensity with
the increase in $x$ and subsequently, the feature around 1.5 eV
becomes intense, narrower and slightly shifted towards higher
binding energies. The decrease in intensity at $\epsilon_F$ is
clearly visible in the symmetrized spectra, $S(\epsilon)$ shown in
Fig. 3(d). Interestingly, $S(\epsilon)$ of $x$~=~0.5 sample exhibits
a soft gap at $\epsilon_F$ and a hard gap appears in $S(\epsilon)$
corresponding to higher $x$. This spectral evolution is remarkably
consistent with the transport properties\cite{kkim}. These results
corresponding to 2-dimensional surface states presumably have strong
implication in realizing recent theoretical
predictions\cite{andreas,dagotto,mohit,paris} and the bulk
properties of this system.

\begin{figure}
\vspace{-2ex}
 \centerline{\epsfysize=4.0in \epsffile{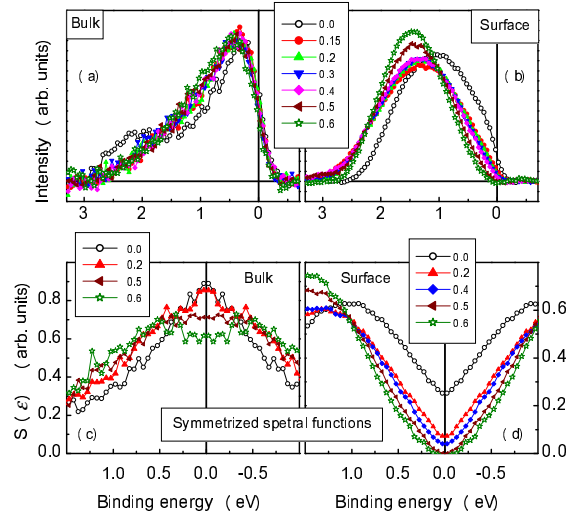}}
%\hspace{2ex}
 \vspace{-16ex}
\caption{(color online) Extracted (a) bulk and (b) surface spectra
of SrRu$_{1-x}$Ti$_x$O$_3$ for various values of $x$. The SDOS
obtained from bulk and surface spectra are shown in (c) and (d),
respectively.}
 \vspace{-2ex}
\end{figure}

\begin{figure}
\vspace{-2ex}
 \centerline{\epsfysize=4.0in \epsffile{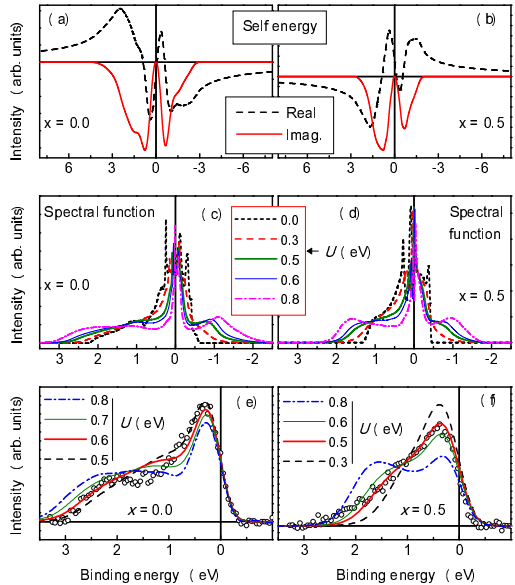}}
%\hspace{2ex}
 \vspace{-10ex}
\caption{(color online) Real and imaginary parts of the self energy
of (a) SrRuO$_3$ and (b) SrRu$_{0.5}$Ti$_{0.5}$O$_3$ obtained by
second order perturbation method following the method of Treglia
{\it et al.}\cite{treglia}. Spectral functions for various values of
$U$ of (c) SrRuO$_3$ and (d) SrRu$_{0.5}$Ti$_{0.5}$O$_3$. Calculated
experimental spectra for different values of $U$ of (e) SrRuO$_3$
and (f) SrRu$_{0.5}$Ti$_{0.5}$O$_3$.}
 \vspace{-2ex}
\end{figure}

The picture is strikingly different in the bulk spectra where the
electronic structure is 3-dimensional. The bulk spectrum of
SrRuO$_3$ exhibits an intense and sharp coherent feature in the
vicinity of $\epsilon_F$ and the incoherent feature appears around
2~eV. The enhancement of $U/W$ due to Ti substitution is expected to
increase the incoherent feature intensity. In sharp contrast, the
intensity of the 2 eV feature reduces significantly and the coherent
feature becomes broad. In addition, the bulk spectra of all the
intermediate compositions appear very similar. The symmetrized bulk
spectra shown in Fig. 3(c) exhibit a small lowering of intensity at
$\epsilon_F$ with the increase in $x$.

Since, $U$ is weak in these highly extended 4$d$
systems\cite{ravi,fujimori}, a perturbative approach may be useful
to understand the role of electron correlation in the spectral
lineshape. We have calculated the bare density of states (DOS) for
SrRuO$_3$ and SrRu$_{0.5}$Ti$_{0.5}$O$_3$ using state-of-the-art
full potential linearized augmented plane wave
method\cite{kmband,wien}. The self energy and spectral functions
were calculated using this $t_{2g}$ partial DOS as done
before\cite{treglia}. The real and imaginary parts of the self
energy are shown in Fig. 4(a) and 4(b), and the spectral functions
for different $U$ values are shown in Fig.~4(c) and 4(d) for
SrRuO$_3$ and SrRu$_{0.5}$Ti$_{0.5}$O$_3$, respectively. The
increase in $U$ leads to a spectral weight transfer outside the LDA
DOS width creating the lower and upper Hubbard bands. Subsequently,
the total width of the LDA DOS diminishes gradually. While these
results exhibit similar scenario as that observed in the most
sophisticated calculations using dynamical mean field theory, the
separation between the lower and upper Hubbard bands is
significantly larger than the corresponding values of $U$. It is
important to note here that the band structure calculations include
the electron-electron interaction term within the local density
approximations. The perturbation calculations in the present case
essentially provide an estimation of the correction in $U$ already
included in the effective single particle Hamiltonian.

In order to compare with the experimental spectra, the calculated
spectral functions are convoluted by Fermi-Dirac distribution
function and the gaussian representing the resolution broadening of
300~meV. The comparison is shown in Figs. 4(e) and 4(f).
Interestingly, the spectral shape corresponding to $U = 0.6 \pm 0.1$
exhibits remarkable representation of the experimental bulk spectra
in both the cases. These results clearly establish that perturbative
approaches and local description of the correlation effects are
sufficient to capture electronic structure of these weakly
correlated systems. The overall narrowing of the valence band
observed in the substituted compounds are essentially a single
particle effect and can be attributed to the reduced degree of
Ru-O-Ru connectivity in these systems. While the high energy scale
features are reproduced remarkably well within this picture, the
occurrence of a {\it pseudogap} at $\epsilon_F$ with increasing $x$
(not visible in Fig.4 due to large energy scale) suggests increasing
role of disorder.

In summary, the high resolution spectra of SrRuO$_3$ exhibit
signature of disorder in the vicinity of the Fermi level.
Introduction of the Ti$^{4+}$ sublattice within the Ru$^{4+}$
sublattice provides a paradigmatic example, where the charge density
near Ti$^{4+}$ sites is close to zero and each Ru$^{4+}$ site
contributes 4 electrons in the valence band. Such large charge
fluctuation leads to a significant change in spectral lineshape and
a dip appears at $\epsilon_F$ ({\em pseudogap}). Interestingly, the
effects are much stronger in the two dimensional (surface)
electronic structure leading to a soft gap at 50\% substitution and
eventually a hard gap appears. Bulk electronic structure
(3-dimensional), however, remains less influenced. A theoretical
understanding of these effects needs consideration of strong
disorder in addition to the electron correlation effects.


\begin{thebibliography}{99}

\bibitem{andreas} A. Fuhrmann, D. Heilmann, and H. Monien, Phys.
Rev. B {\bf 73}, 245118 (2006).

\bibitem{dagotto} S.S. Kancharla and E. Dagotto, Phys. Rev. Lett.
{\bf 98}, 016402 (2007).

\bibitem{mohit} Arti Garg, H.R. Krishnamurthy, and Mohit Randeria,
Phys. Rev. Lett. {\bf 97}, 046403 (2006).

\bibitem{paris} N. Paris, K. Bouadim, F. Hebert, G.G. Batrouni, and R.T.
Scalettar, Phys. Rev. Lett. {\bf 98}, 046403 (2007).

\bibitem{jkim} J. Kim, J.-Y. Kim, B.-G. Park, and S.-J. Oh, Phys. Rev. B
{\bf 73}, 235109 (2006), M. Abbate, J.A. Guevara, S.L. Cuffini, Y.P.
Mascarenhas, and E. Morikawa, Eur. Phys. J. B {\bf 25}, 203 (2002).

\bibitem{dd} S. Ray, D.D. Sarma, and R. Vijayaraghavan, Phys. Rev. B
{\bf 73}, 165105 (2006).

\bibitem{kkim} K.W. Kim, J.S. Lee, T.W. Noh, S.R. Lee, and K. Char,
Phys. Rev. B {\bf 71}, 125104 (2005).

\bibitem{sscO1s} R.S. Singh and K. Maiti, Solid State Commun, {\bf
140}, 188 (2006).

\bibitem{cao} G. Cao, S. McCall, M. Shepard, J.E. Crow, and R.P. Guertin,
Phys. Rev. B {\bf 56}, 321 (1997).

\bibitem{ravi} K. Maiti and R.S. Singh, Phys. Rev. B {\bf 71},
161102(R) (2005).

\bibitem{kmband} K. Maiti, Phys. Rev. B {\bf 73}, 235110 (2006).

\bibitem{fujimori} M. Takizawa, D. Toyota, H. Wadati, A. Chikamatsu,
H. Kumigashira, A. Fujimori, M. Oshima, Z. Fang, M. Lippmaa, M.
Kawasaki, and H. Koinuma, Phys. Rev. B {\bf 72}, 060404(R) (2005).

\bibitem{kmepl} K. Maiti, R.S. Singh, and V.R.R. Medicherla,
Europhys. Lett. (in print); Condmat/0604648.

\bibitem{altshuler-aronov} B.L. Altshuler and A.G. Aronov, Solid
State Commun. {\bf 30}, 115 (1979).

\bibitem{ddsprl} D.D. Sarma {\it et al.}, Phys. Rev. Lett. {\bf
80}, 4004 (1998).

\bibitem{efros} A.L. Efros and B.I. Shklovskii, J. Phys. C: Solid
State Phys. {\bf 8}, L49 (1975).

\bibitem{massey} J.G. Massey and M. Lee, Phys. Rev. Lett. {\bf
75}, 4266 (1995).

\bibitem{wien} P. Blaha, K. Schwarz, G.K.H. Madsen, D. Kvasnicka, and J. Luitz,
WIEN2k, An Augmented Plane Wave + Local Orbitals Program for
Calculating Crystal Properties (Karlheinz Schwarz, Techn.
Universit\"{a}t Wien, Austria), 2001. ISBN 3-9501031-1-2.

\bibitem{treglia} G. Treglia {\it et. al.}, J. Physique {\bf 41}, 281 (1980);
{\it ibid}, Phys. Rev. B {\bf 21}, 3729 (1980); D.D. Sarma {\it et
al.}, Phys. Rev. Lett. {\bf 57}, 2215 (1986).

\end{thebibliography}
\end{document}